# Binding of the immunomodulatory drug Bz-423 to mitochondrial F$_o$F$_1$-ATP synthase in living cells by FRET acceptor photobleaching

Ilka Starke[a,b], Kathryn M. Johnson[c], Jan Petersen[b,d], Peter Gräber[b], Anthony W. Opipari, Jr.[e], Gary D. Glick[c], Michael Börsch*[a,f,g]

[a] Single Molecule Microscopy Group, Jena University Hospital, Friedrich Schiller University, Nonnenplan 2 - 4, 07743 Jena, Germany; [b] Institute for Physical Chemistry, Albert-Ludwigs-University, Albertstrasse 23a, 79104 Freiburg, Germany; [c] Department of Chemistry, University of Michigan, 930 N University Ave, Ann Arbor, MI 48109-1055, USA; [d] Department of Biochemistry and Molecular Biology, Faculty of Medicine, Nursing and Health Sciences, Monash University, Clayton, Victoria 3800, Australia; [e] Department of Obstetrics and Gynecology, University of Michigan, 930 N University Ave, Ann Arbor, MI 48109-1055, USA; [f] Center for Medical Optics and Photonics (CeMOP), Jena, Germany; [g] Abbe Center of Photonics (ACP), Jena, Germany

## ABSTRACT

Bz-423 is a promising new drug for treatment of autoimmune diseases. This small molecule binds to subunit OSCP of the mitochondrial enzyme F$_o$F$_1$-ATP synthase and modulates its catalytic activities. We investigate the binding of Bz-423 to mitochondria in living cells and how subunit rotation in F$_o$F$_1$-ATP synthase, i.e. the mechanochemical mechanism of this enzyme, is affected by Bz-423. Therefore, the enzyme was marked selectively by genetic fusion with the fluorescent protein EGFP to the C terminus of subunit γ. Imaging the threedimensional arrangement of mitochondria in living yeast cells was possible at superresolution using structured illumination microscopy, SIM. We measured uptake and binding of a Cy5-labeled Bz-423 derivative to mitochondrial F$_o$F$_1$-ATP synthase in living yeast cells using FRET acceptor photobleaching microscopy. Our data confirmed the binding of Cy5-labeled Bz-423 to the top of the F$_1$ domain of the enzyme in mitochondria of living *Saccharomyces cerevisiae* cells.

## 1. INTRODUCTION

Förster resonance energy transfer (FRET) is a powerful method to determine distances between two molecular targets, either in purified solution or in complex environments like living cells. FRET acts as a precise ruler for measurements within a distance range of 2 to 10 nm and with sub-nanometer resolution[1]. FRET microscopy can provide spatial and temporal information of molecular interactions, i.e. on the sub-cellular or sub-micrometer level, respectively. For example, FRET can be localized within specific organelles of the cell, in lipid membranes or in nucleic acid environments. Intensity-based FRET is detected as in an apparent loss of FRET donor intensity and simultaneous gain of FRET acceptor intensity, or a relative shift of intensities from the FRET donor fluorophore to FRET acceptor fluorophore. FRET is also detectable by changes of the FRET donor fluorescence lifetime, because energy transfer shortens the FRET donor lifetime. Alternatively, FRET can be revealed by selectively photobleaching the FRET acceptor dye and by the subsequent increase in FRET donor fluorescence. Here, FRET acceptor photobleaching was applied to monitor binding of a fluorescent derivative of Bz-423 to mitochondrial F$_o$F$_1$-ATP synthase.

.................................................................................................................................................................

* michael.boersch@med.uni-jena.de; phone +49 36419396618; fax +49 3641933750; http://www.m-boersch.org

To reveal a binding site for small molecules like the immunomodulatory drug Bz-423[2-27] within a living cell, both the target and the drug have to be labeled with different fluorophores for FRET. The 1,4-benzodiazepine Bz-423 was modified at the N1 position of the heterocycle by a flexible alkyl chain linker instead of a methyl group. The accessible amino group of the linker provided the functionality for specific reaction with a NHS ester derivative of the red-fluorescent dye Cyanine 5 (Cy5). Compound Bz-423-Cy5 is the FRET acceptor in our measurements. The structure of the molecule is shown in Figure 1a.

Fusion of fluorescent proteins to the mitochondrial target of Bz-423, i.e. $mF_oF_1$-ATP synthase, have been reported previously[28-30]. Different subunits of the enzyme from *Saccharomyces cerevisiae* could be tagged with green fluorescent protein GFP, including the 'oligomycin sensitivity conferral protein' subunit OSCP or the rotary γ-subunit of the $F_1$ domain[30]. Here we applied a similar strategy to label $mF_oF_1$-ATP synthase in living yeast cells with the FRET donor fluorophore. Briefly, the C terminus of the γ-subunit was prolonged with a peptide linker and a yeast-enhanced GFP was fused to this peptide linker (green box in Figure 1b).

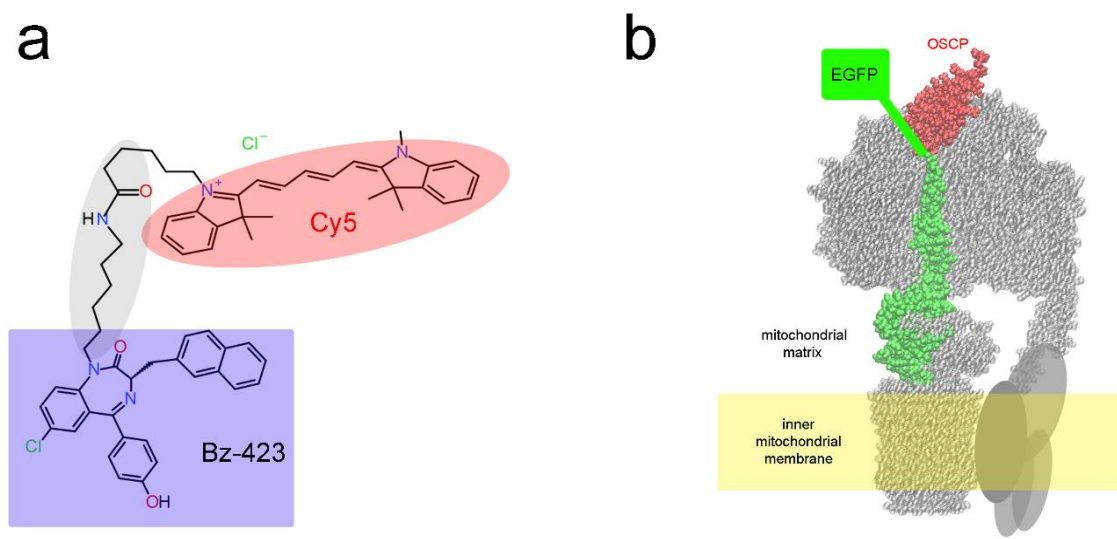

**Figure 1**. **a**, structure of Bz-423-Cy5 with the distinct moieties of Bz-423 (blue box), the aminohexyl linker (grey ellipse) and Cy5 (red ellipse). **b**, model of mitochondrial $F_oF_1$-ATP synthase with atomic structural details according to PDB 4B2Q[31]. Highlighted are subunit γ (green dots, center) and OSCP (red dots, top). The fluorescent protein EGFP (green box) is fused to the C terminus of subunit γ *via* a peptide linker (green stick). The yellowish bar indicates the lipid bilayer level of the inner mitochondrial membrane.

Fusion of fluorescent proteins with a length of ~4.2 nm and a diameter of ~2.5 nm (and a mass of ~27 kDa) to subunits of enzymes might affect their function and their conformational dynamics but also protein expression level. $F_oF_1$-ATP synthase is the ubiquitous enzyme that operates as a proton- (or $Na^+$)-driven rotary motor in bacterial membranes, in the chloroplast and in the inner mitochondrial membranes[32, 33]. Central subunits γ, ε, and c rotate stepwise within the enzyme to synchronize and enable the chemical synthesis of adenosine triphosphate (ATP) in three catalytic binding sites. Since nearly 20 years, a focus of our research has been the unravelling of conformational dynamics of the rotary motor $F_oF_1$-ATP synthase of *Escherichia coli* by single-molecule FRET[34-80]. FRET changes between two specifically attached fluorophores - one on a rotary subunit and another one on a static position or subunit - in single enzymes were recorded in a confocal microscope equipped with avalanche photodiodes for time-correlated single photon counting. For some measurements we have used an EGFP fusion to the static subunit *a* as FRET donor[49, 54, 55, 65, 66]. The purified EGFP mutant enzyme exhibited similar ATP synthesis rates after reconstitution into liposomes *in vitro* so that we considered this fusion mutant as functional. However, according to qualitative SDS-PAGE analysis the relative amount of $F_oF_1$-ATP synthases with a mutated fusion to subunit *a* was always significantly lowered in the isolated membrane protein fraction compared to a 'wildtype'-like enzyme without this large additional mass[74, 76].

For a critical assessment of the EGFP fusion to $mF_oF_1$-ATP synthase and as an indicator for significant perturbation of catalytic function, measurements of yeast cell growth rates under conditions for forced oxidative phosphorylation or microscopic imaging of morphology changes of the mitochondria could be used. Previously, deletion of subunits of $mF_oF_1$-ATP synthase as well as aging processes resulted in swelling of mitochondria, shortening and rearrangement of the cristae[81]. Therefore, we evaluated the use of optical superresolution imaging of living *S. cerevisiae* cells by structured illumination microscopy (SIM). We show that yeast mitochondria with EGFP fusion to $mF_oF_1$-ATP synthase possess the well-known tubular network structures around the vacuoles in the presence of ethanol and succinate in minimal media for forced oxidative phosphorylation, but change shape to fragmented small mitochondria in the absence of oxygen.

Bz-423 was shown previously to bind to different sites on OSCP of $mF_oF_1$-ATP synthase in sub-mitochondrial particles from bovine heart[7, 10, 20]. The binding constants were found in the low micromolar range. Our aim was to evaluate the use of living *Saccharomyces cerevisiae* cells for imaging Bz-423 binding. Here we provide evidence that a labeled Bz-423 derivative is detected in the mitochondria and is binding to the top of the $F_1$ part of $mF_oF_1$-ATP synthase, i.e. near the C terminus of the γ-subunit and, therefore, likely on OSCP of the yeast enzyme.

## 2. EXPERIMENTAL PROCEDURES

### 2.1 Bz-423-Cy5 and EGFP-fusion to mitochondrial $F_oF_1$-ATP synthase in *Saccharomyces cerevisiae*

*Synthesis of Bz-423-Cy5* - The Bz-423-Cy5 conjugate was synthesized by condensing Cy5-NHS with N1-hexylamino-Bz-423 analog as previously described[5] (K. M. Johnson, "*Elucidation of the Target and Molecular Mechanism of Action of the Immunomodulatory Benzodiazepine, Bz-423*". PhD Thesis, University of Michigan, 2005). The structure of the compound Bz-423-Cy5 is shown in Figure 1a.

*Staining mitochondria in S. cerevisiae with Bz-423-Cy5* - To detect binding of Bz-423-Cy5 to mitochondrial $F_oF_1$-ATP synthase in living *S. cerevisiae* cells, we used the following mutant. The enhanced green fluorescent protein, EGFP, was linked to the C terminus of the γ-subunit and a cysteine was mutated into subunit *b* (details to be published elsewhere, J. Petersen, I. Starke and P. Gräber). A simple model of the $mF_oF_1$-ATP synthase is shown in Figure 1b. The double mutant showed ATP synthesis activity, i.e. catalytic rates were determined *in vitro* after protein purification and reconstitution of $mF_oF_1$-ATP synthase to liposomes[82, 83].

Cells were grown in 10 ml YPD medium (yeast extract 1 %, peptone 1 %, glucose 2 %) at 28°C for 25 h with gentle shaking at 220 rpm. Cells were harvested by centrifugation and washed with buffer A (20 mM tricine, 20 mM succinate, 0.6 mM KCl, pH 8.0) in the presence of 2% EtOH or 2% glucose, respectively. Staining mitochondria was achieved by incubating the cells with 4 μm Bz-423-Cy5 in buffer A for 8 h at 28°C with gentle shaking at 220 rpm. Afterwards, cells were washed twice with 1 ml buffer A and kept in buffer A in the presence of 2% EtOH (or 2% glucose) at 28° until imaging on the Nikon microscope.

### 2.2 Structured Illumination Microscopy (SIM)

The Nikon N-SIM microscope was based on an inverted software-controlled Nikon Ti Eclipse with two independent fluorescence filter wheels stacked on top of each other for dual use. Superresolution microscopy with N-SIM comprised four continuous-wave lasers (Coherent, 405 nm, 488 nm, 561 nm and 640 nm) coupled either by a multi-mode fiber or a single-mode fiber to one of four different gratings for 3D-SIM, 2D-SIM or TIRF-SIM mode. Images were recorded by a cooled Andor EMCCD camera (iXon DU-897) using the Nikon FITC optical filter set for SIM. A fast autofocus system maintained spatial stability of the microscopic samples against drift in z-direction. We used a 100x Apo TIRF oil objective with N.A. 1.49 for SIM. An additional 2.5x-magnifying lens system before the EMCCD resulted in an effective pixel size of 64 nm.

*S. cerevisiae* cells were imaged in a home-built sample chamber using cell immobilization on a thin agarose film as described previously[84]. Briefly, glass slides were oxygen plasma-treated for 5 minutes to smoothen the surface and to remove fluorescent impurities. Subsequently, 50 μl of a hot low melting agarose solution in purified $H_2O$ (Millipore) were placed on the glass slide that was heated to 30°C. The agarose gel droplet was deducted with a second glass slide. After 30 min drying time, yeast cells were placed as a 5 μl suspension on the agarose film, and samples were sealed with a plasma-treated cover slide. SIM images were recorded at 22°C. Nikon analysis software was used for SIM image reconstruction and 3D visualization.

### 2.3 FRET microscopy

The inverted Nikon N-SIM / N-STORM superresolution microscope was used for FRET microscopy using the N-STORM laser excitation pathway in regular widefield excitation mode[76]. Cells were imaged using 488 nm laser excitation (Agilent MCL400B system for N-STORM) at a low power level of "10%" to minimize photobleaching. Images were recorded with a 60x water immersion objective (PlanApo IR, N.A. 1.27). Fluorescence detection between 500 and 550 nm was achieved by using an Andor iXon DU-897 EMCCD camera. The pixel size was 267 nm without an additional optical magnification lens before the EMCCD. Yeast cells were concentrated by low speed centrifugation and immobilized as a suspension, i.e. in a 10 µl droplet, on a small petri dish (Mobitec µdish with grid) with 170 µm cover glass bottom for microscopy. After 5 min, cells had settled down to the cover glass. Before imaging, a 50 µl droplet of buffer A with 2% EtOH was added on top of the cells. A set of 10 subsequent images with 16 ms integration time for each condition were recorded before the laser was turned off. Afterwards the same field of view was illuminated with 640 nm at high power (power level "50%", Agilent MCL400B system) for 20 sec per cycle to selectively photobleach the Cy5 fluorophore on Bz-423. Subsequently, another set of 10 images was recorded with identical laser power at 488 nm using the same EMCCD parameters as before.

### 2.4 FRET data analysis

Quantitative analysis of the FRET data stored as image stacks in Nikon-specific 'nd2 format' was achieved using the free imaging software ImageJ (version 1.49v or the Fiji version thereof). Individual cells were selected manually in each set of images by drawing a circular region of interest (ROI) around one cell. After determining minimum and maximum pixel intensities in each ROI per frame, the max-min intensity difference was calculated (OriginPro 8G) and averaged over the set of 10 ROI images. The same cell was selected after laser exposure with 640 nm, and averaged max-min differences were calculated again. As a result of strong illumination with 640 nm, both minimum and maximum intensities in the ROI, or a single cell, respectively, increased for cells stained with Bz-423-Cy5.

## 3. RESULTS

### 3.1 Imaging mitochondria in living yeast cells with N-SIM

$F_oF_1$-ATP synthases of the yeast *Saccharomyces cerevisiae* were mutated to carry a fluorescent protein as fusion to the C terminus of the γ-subunit. This enzyme is often called complex V of the set of five large protein machines in the inner mitochondrial membrane that couple redox chemistry, electron and proton transport reactions for the production of a proton-motive force, PMF. The PMF across the inner mitochondrial membrane is required for the synthesis of ATP by $F_oF_1$-ATP synthase. Electron microscopic images at cryogenic temperatures showed that the distribution of $F_oF_1$-ATP synthase within the inner mitochondrial membrane is likely non-homogeneous. In yeast and other eukaryotic organisms, the enzymes are located mainly at the rim of the cristae, and are orientated and aligned like a double-row of enzymes[85].

Because loss of $F_oF_1$-ATP synthase subunits by genetic manipulation resulted in altered morphologies of mitochondria, we asked if our mutated γ-subunit of the enzyme might affect the dynamic processes of mitochondrial shape changes by fusion and fission. Mitochondria are small organelles with less than 500 nm in diameter and a few µm length. Therefore, optical superresolution imaging beyond the diffraction limit is required to identify subtle changes. SIM allows to double the optical resolution down to a ~110 nm limit.

We used a Nikon N-SIM superresolution microscope to image yeast mitochondria in living cells. Cells were grown in the presence of EtOH and succinate to enforce oxidative phosphorylation. A droplet of the cell suspension was placed onto a agarose-coated glass cover slide to immobilize the cells. The sample chamber was sealed with nail polish. Cells were imaged immediately at 22°C using 488 nm excitation and the Nikon 3D-SIM mode with a 100x TIRF objective, oil immersion, N.A. 1.49. To visualize the three-dimensional orientation of mitochondria, z-stacks of 15 to 25 image planes were recorded. Figure 2 shows reconstructed SIM images of EGFP-tagged FoF1-ATP synthases in mitochondria of yeast cells in the presence of EtOH and succinate, at the beginning of the measurements (Figure 2a) and after about 1 h (Figure 2b).

Immediately after embedding the cells on agarose, mitochondria appeared as a three-dimensional tubular network located at the periphery of the cells with 5 to 7 µm in diameter. In Figure 2a, recording a stack of 15 subsequent image planes with 500 nm spacing covered the larger cell almost completely. SIM reconstruction yielded doubled resolution and, accordingly, a pixel size of 32 nm in the reconstructed SIM images. The full-width-half-maximum values of intensity histograms were measured perpendicular to the long axis of mitochondrial and revealed a mitochondrial diameter of 250 to 300 nm. The lengths of mitochondria were in the range of several µm.

However, the filamentous structures of mitochondria changed over time. The dynamic processes of mitochondrial fission started stochastically, and cells with both tubular and fragmented mitochondria simultaneously were found. After about 1 h, all cells exhibited fragmented mitochondria as shown in Figure 2b. The mitochondria were located at the periphery of the yeast cells. The droplet-like mitochondria had diameters of about 250 to 500 nm, and the apparent number of fragmented mitochondria was high. The z-stack consisted of 25 image planes with 300 nm spacing in Figure 2b. Because each SIM image for one image plane required a 2-second recording of 15 sub-images, z-spacing of the image planes was chosen to complete a z-stack within less than 1 min.

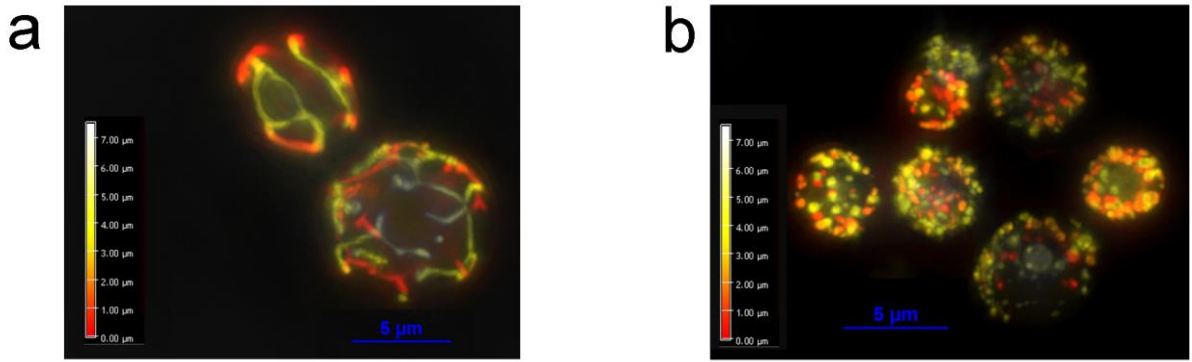

**Figure 2**. SIM images of EGFP-tagged $F_oF_1$-ATP synthases in mitochondria of living *S. cerevisiae* at 22°C with false color-coded z position. Red is 0 µm above cover glass, yellow is ~ 3 to 4 µm above cover glass, white is ~ 7 µm above cover glass. Images represent z-stacks in depth-coded maximum intensity projection (Nikon NIS Elements software). Imaging buffer contained 20 mM succinate and 2% EtOH. Blue scale bar is 5 µm, the pixel size is 32 nm after SIM reconstruction. **a**, Elongated filamentous mitochondria, **b**, fragmented mitochondria.

### 3.2 FRET Imaging

Binding of Bz-423-Cy5 to $mF_oF_1$-ATP synthase in living yeast was detected using Förster resonance energy transfer. FRET donor was the fluorescent protein EGFP fused to the γ-subunit of the mitochondrial enzyme. Cy5 on Bz-423 acted as FRET acceptor. To avoid mitochondrial oxygen depletion during FRET imaging, yeast cells were placed as a droplet on an open petri dish with 170 µm cover glass suitable for microscopy with high N.A. objectives. The coating of the cover glass supported an immobilization of yeast cells after settling down to the glass surface. However, some cells still moved due to Brownian motion.

FRET imaging by acceptor photobleaching is a multi-step process. The first widefield fluorescence image with excitation at 488 nm yields the reference intensity distribution of the FRET donor in the presence of the FRET acceptor. Therefore, we aimed at minimal excitation power and short exposure times. The laser power at 488 nm was set to "10%" and a set of 10 subsequent images with 16 ms integration time was recorded. The resulting FRET donor intensities in each image were low (always less than 700 a.u. of 16384 at 14 bit resolution). Averaging the 10 sub-images yielded the FRET images shown in Figure 3. To enable fast image recording, the field-of-view on the EMCCD camera chip was reduced to 256x256 pixels. Using a 60x water immersion objective (Nikon, N.A. 1.27) and avoiding an additional magnification lens before the EMCCD resulted in a pixel size of 267 nm, i.e. ten times larger than in the SIM recording above. With this overall microscope magnification, many yeast cells could be recorded simultaneously and reliable sample statistics could be achieved. Comparing the averaged images in Figures 3a-c recorded immediately one after another revealed that the 488 nm laser power did not induce a measurable photobleaching of the FRET donor.

The second step for FRET imaging requires photobleaching of the FRET acceptor. Here we used the high power, 640 nm laser of the Nikon Agilent laser box and induced photobleaching of Cy5 with "50%" of laser power for 20 seconds. During the photobleaching step the EMCCD camera does not record data. After 20 s, laser excitation is switched back to 488 nm with "10%" laser power. Figures 3d and e show the false-colored fluorescence images of the FRET donor of the same yeast cells. A strong increase in green and yellow-coded pixels in many cells indicate that FRET donor intensity increased. In some cells, mitochondria could be identified as bright spots at the periphery of the cell. However, given the resolution of the FRET images with 267 nm pixel size, we were unable to discriminate and resolve filamentous mitochondria.

We repeated the 20-second photobleaching step with 640 nm two times. Figure 3f shows FRET donor intensities after the second bleaching, Figure 3g after the third bleaching. In general, FRET donor intensities increased after the second and again after the third bleaching. However, cell walls and internal mitochondrial structures appeared to be "de-focussed" in Figures 3f and 3g. A slow drift of the microscope stage in z-direction could be the reason because we required a total measurement time of about 4 minutes from Figures 3a to 3g.

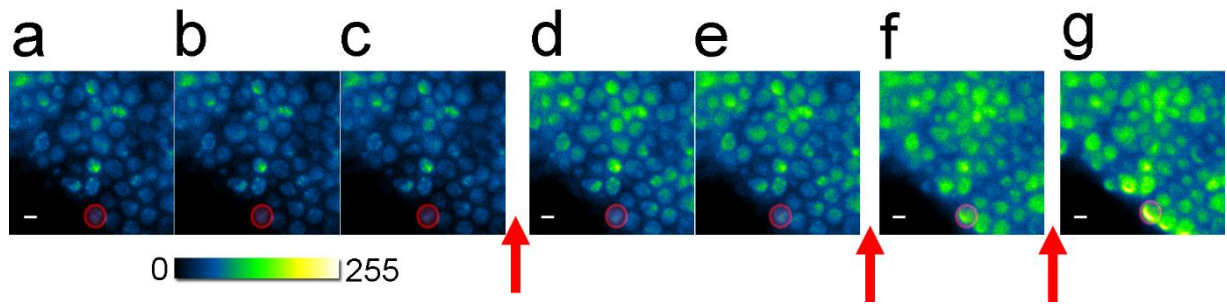

**Figure 3**. Time series of fluorescence images of EGFP-tagged $F_oF_1$-ATP synthases in mitochondria of living *S. cerevisiae*. Yeast cells were stained for 8 h with 4 μm Bz-423-Cy5. Excitation with 488 nm at low laser intensity ("10%" Agilent MCL400B) using a 60x water immersion objective. Fluorescence intensities were rescaled to the same minimal (100 a.u.) and maximal values (750 a.u.) and color-coded to 8 bit. Red arrows indicate each 20-second photobleaching step with 640 nm at high laser power ("50%" Agilent MCL400B). White scale bar is 5 μm, the pixel size is 267 nm. Red circles highlight the analyzed single yeast cell in all images. **a - c**, consecutive time series of fluorescence image of the same yeast cell population. Shown is the average of ten images. **d, e,** consecutive series of fluorescence images immediately after the first 20 second high power exposure with 640 nm. **f**, fluorescence image immediately after the second 20 second high power exposure with 640 nm. **g**, fluorescence image immediately after the third 20 second high power exposure with 640 nm.

### 3.3 FRET analysis of a single cell

To measure the increase of FRET donor fluorescence quantitatively after photobleaching the Cy5 moiety on Bz-423, we analyzed individual cells. A cell was selected by manually drawing a circular region-of-interest (ROI) filter with 25 pixel in diameter, or 6.7 μm, respectively. The ROI was chosen larger than the bright parts of the cell in order to include also some pixel information from outside the cell, i.e. as a control for the background intensity. The ROI position of the cell was controlled in each of the 10 sub-images and was adjusted if necessary. ROI positions could change due to Brownian motion of cells and due to thermal drift of the microscope stage within the several minutes of recording. In each ROI, intensity analysis was achieved using ImageJ software. We extracted minimum and maximum pixel intensities of each ROI, averaged minimum and maximum intensities over the ten sub-images, and calculated the averaged max-min differences. These numbers for a single cell are plotted in Figure 4 (the selected cell is indicated by the highlighted red circle in Figure 3).

In the first three subsequent series of FRET imaging, the minimum and maximum pixel intensities varied slightly, but a loss of intensity with exposure time was not detected (Figures 4 a - c). Small fluctuations of both minimum and maximum intensities were found for all selected cells (data not shown) and were attributed to Poisson noise for low intensities.

After the first photobleaching step the max-min intensity difference (blue bars in Figure 4) increased from 265±8 a.u. to 329±9 a.u., i.e. corresponding to an increase of 24%. After the second bleaching step, the max-min intensity difference was 415 a.u. (57%), and after the third bleaching step 538 a.u. (103%). This selected cell showed a prominent change of average FRET donor intensity; however, other cells exhibited a similar fluorescence increase after photobleaching.

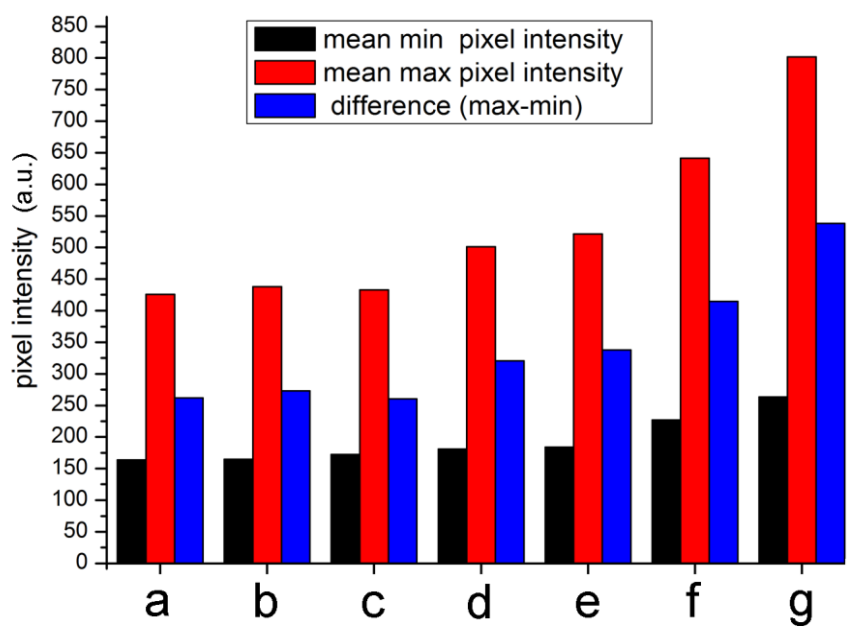

**Figure 4**: Pixel intensities of the single yeast cell highlighted by a red circle in Figure 3. Minimal (black bars) and maximal (red bars) pixel intensities as well as max-min differences (blue bars) after averaging the set of ten individual sub-images per time series. Excitation with 488 nm. **a - c**, pixel intensities before the first high power exposure with 640 nm. **d, e,** pixel intensities after the first 20-second 640 nm exposure, **f,** after the second, and **g,** after the third high power photobleaching step with 640 nm.

## 4. DISCUSSION

Bz-423 is a 1,4-benzodiazepine molecule that selectively kills pathogenic lymphocytes which cause autoimmune diseases like systemic lupus erythematosus. The drug induces apoptosis *via* increased $O_2^-$-production in coupled, actively respiring mitochondria. The subcellular target was identified by *in vitro* experiments with isolated mitochondria and with sub-mitochondrial particles. Accordingly Bz-423 binds to subunit OSCP of mitochondrial $F_oF_1$-ATP synthase. At concentrations between 5 to 10 μm Bz-423 reduced both ATP hydrolysis and ATP synthesis rates to ~ 50%. Similar concentrations were found as $EC_{50}$ in living cells.

Our aim was to directly demonstrate binding of Bz-423 to $mF_oF_1$-ATP synthase in living cells. We chose the yeast *S. cerevisiae* as a model organism with a mitochondrial enzyme which could be genetically modified. A fusion of EGFP to the C terminus of the γ-subunit was used as a marker of the enzyme. The EGFP fluorophore was expected to be located near the binding sites of Bz-423 on subunit OSCP. The distance between EGFP on the γ-subunit and Cy5 linked to Bz-423 in a possible binding site on OSCP is not known yet. However, a simple structural model for the enzyme could be obtained. The expected distances should be in the range of 5 to 10 nm.

This distance range matches the maximum sensitivity for FRET for the fluorophores EGFP and Cy5. In principle, FRET will be accompanied by FRET acceptor fluorescence when we excite the FRET donor at 488 nm and energy is transferred to the acceptor. However, we could not resolve intracellular structures like mitochondria using the FRET acceptor fluorescence of Cy5. Because the binding constants of Bz-423-Cy5 were likely in the micromolar range, we had applied a 4 μm Bz-423-Cy5 solution for staining the yeast cells. After washing the cells with buffer, the cells maintained a bluish appearance, which indicated a successful uptake of Bz-423-Cy5 and a high remaining intracellular concentration of the FRET acceptor. Direct excitation of the FRET acceptor with 640 nm at very low power was also not possible. The fluorescence of Cy5 was still too high for regular imaging with an EMCCD camera and did not show any structural details inside the yeast cells (data not shown).

Instead, FRET between Bz-423-Cy5 and $mF_oF_1$-ATP synthase could be revealed analyzing FRET donor fluorescence increase after acceptor photobleaching. Using minimal excitation power at 488 nm and short exposure of 10 times 17 ms avoided untimely photobleaching of the FRET donor. A 20-second high power exposure with 640 nm was sufficient to partly photobleach Bz-423-Cy5. Subsequent fluorescence imaging resulted in a ~20% increase in background-corrected FRET donor intensities. These findings supported our assumption that the Cy5-labeled immunomodulatory drug Bz-423 also binds to the OSCP subunit of $F_oF_1$-ATP synthase in mitochondria of living yeast cells. In the controls (data not shown) with un-stained yeast cells we did not observe an increase of EGFP fluorescence upon high power irradiation with 640 nm. A quantitative FRET analysis with many analyzed cells and more controls will be published elsewhere.

We will expand this FRET-based Bz-423 binding analysis to human cells as the real target. It was shown previously that the C terminus of the γ-subunit can be tagged with a fluorescent protein without affecting mitochondrial morphology or $F_oF_1$-ATP synthase function in human cells[86, 87]. In addition to intensity-based FRET analysis, donor fluorescence lifetime imaging (FLIM) could help to determine FRET efficiencies and to estimate the distances between the fluorophore on the γ-subunit and the Bz-423 binding sites on OSCP. Furthermore, superresolution imaging like STED imaging of mitochondria might be useful to quantify Bz-423 action and to understand how this molecule affects the catalytic function of $F_oF_1$-ATP synthase at work.

**Acknowledgements**

We gratefully acknowledge the collaboration with Mark Prescott (Monash University) to generate γ-subunit mutants in mitochondrial enzyme of *Saccharomyces cerevisiae*. This work was supported by NIH grant AI-47450 (to G.D.G.) and in part by DFG grant BO 1891/15-1 (to M.B.) The Nikon N-SIM / N-STORM microscope was funded by the State of Thuringa (grant FKZ 12026-515 to M.B.).